\documentclass[cameraready]{Interspeech}

\title{Mind the Gap: Detecting Cluster Exits for Robust Local Density-Based\texorpdfstring{\\}{ }Score Normalization in Anomalous Sound Detection}

\author[affiliation={1,2}, orcid=0000-0003-4200-9129, correspondingauthor]{Kevin}{Wilkinghoff}
\author[affiliation={3}, orcid=0000-0002-8597-6795]{Gordon}{Wichern}
\author[affiliation={3}, orcid=0000-0002-3451-171X]{Jonathan}{Le Roux}
\author[affiliation={1,2}, orcid=0000-0001-6856-8928]{Zheng-Hua}{Tan}

\address{
    $^1$ Department of Electronic Systems, Aalborg University, Aalborg, Denmark \\
    $^2$ Pioneer Centre for Artificial Intelligence, Copenhagen, Denmark \\
    $^3$ Mitsubishi Electric Research Laboratories (MERL), Cambridge, MA, USA
}

\email{kevin.wilkinghoff@ieee.org, wichern@merl.com, leroux@merl.com, zt@es.aau.dk}


\usepackage{cite}
\usepackage[capitalize,nameinlink]{cleveref}

\crefname{figure}{Fig.\@}{Figs.\@}
\Crefname{figure}{Figure\@}{Figures\@}

\crefname{table}{Table\@}{Tables\@}
\Crefname{table}{Table\@}{Tables\@}

\usepackage{csquotes}
\usepackage{acronym}
\usepackage{tikz,pgfplots}
\usepgfplotslibrary{groupplots,units}
\usetikzlibrary{patterns}
\pgfplotsset{compat=newest}
\usetikzlibrary{shapes.geometric, arrows, calc, fit, shapes, backgrounds}
\usepgfplotslibrary{colorbrewer}

\acrodef{act}[ACT]{auxiliary classification task}
\acrodef{asd}[ASD]{anomalous sound detection}
\acrodef{knn}[k-NN]{k-nearest neighbors}
\acrodef{ldn}[LDN]{local density-based anomaly score normalization}
\acrodef{ced}[CED]{cluster-exit detection}
\acrodef{varmin}[VarMin]{variance minimization}

\newcommand{\score}{\mathcal{A}}
\newcommand{\dist}{\mathcal{D}}

\usepackage{adjustbox}
\usepackage{nicematrix}
\usepackage{booktabs}
\usepackage{multirow}
\usepackage{threeparttable}

\begin{document}

\maketitle

\begin{abstract}
Local density-based score normalization is an effective component of distance-based embedding methods for anomalous sound detection, particularly when data densities vary across conditions or domains. In practice, however, performance depends strongly on neighborhood size. Increasing it can degrade detection accuracy when neighborhood expansion crosses cluster boundaries, violating the locality assumption of local density estimation. This observation motivates adapting the neighborhood size based on locality preservation rather than fixing it in advance. We realize this by proposing cluster exit detection, a lightweight mechanism that identifies distance discontinuities and selects neighborhood sizes accordingly. Experiments across multiple embedding models and datasets show improved robustness to neighborhood-size selection and consistent performance gains.
\end{abstract}

\acresetall

\section{Introduction}
Semi-supervised \ac{asd} for machine condition monitoring aims to detect anomalous machine sounds using training data that contain only recordings of normal operation. In real-world deployment, however, systems often encounter domain shifts, such as changes in operating conditions, background noise, recording devices, or environments, which alter the sound distribution even though the machine remains in a normal state. This setting is formalized in recent DCASE challenges \cite{dohi2022description,dohi2023description,nishida2024description,nishida2025description}, where models are trained with many normal samples from a source domain and only a few from a target domain, yet are expected to perform equally well in both domains at test time.

To handle this setting, many state-of-the-art \ac{asd} systems operate in learned embedding spaces using distance-based scores or density estimates. Each audio recording, typically several seconds long, is mapped to a fixed-dimensional embedding vector by a neural network, either trained with task-specific surrogate objectives for machine sounds \cite{giri2020self,primus2020anomalous,wilkinghoff2021sub-cluster,venkatesh2022improved,jiang2024anopatch,jiang2025adaptive} or pre-trained on large-scale audio datasets \cite{mezza2023zero-shot,saengthong2024deep,wu2025towards,han2025exploring,zhang2025echo,fan2025fisher,wilkinghoff2026temporal}. Anomalies are identified by comparing embeddings of test samples to those of normal training data in the embedding space.

Recently, \ac{ldn} \cite{wilkinghoff2025keeping,wilkinghoff2025local} has been proposed to improve robustness to domain shifts in this embedding-based setting \cite{wilkinghoff2025handling}. \Ac{ldn} normalizes distance-based anomaly scores using statistics from a local neighborhood in the embedding space, compensating for variations in local data density. This is particularly important when detection relies on a single, domain-agnostic threshold despite highly non-uniform reference densities. However, \ac{ldn} introduces the neighborhood size as a critical hyperparameter.

Empirically, increasing the neighborhood size beyond one or two neighbors often leads to systematic performance degradation. While larger neighborhoods are generally expected to yield more stable density estimates, existing approaches typically fix the neighborhood size to a small value \cite{wilkinghoff2025local,fujimura2025asdkit}, offering limited insight into why performance deteriorates as the neighborhood expands. \Ac{varmin} \cite{matsumoto2025adjusting} can partially mitigate this effect, but the fundamental sensitivity to neighborhood size remains unresolved.

In this work, we show that the observed degradation is not caused by larger neighborhoods per se, but occurs when neighborhood expansion crosses cluster boundaries in the embedding space. Once neighbors outside the local region are included, the locality assumption underlying \ac{ldn} is violated, corrupting local density estimates and destabilizing score normalization. This insight motivates a general design principle: Neighborhood sizes should adapt based on whether locality is preserved, rather than being fixed a priori. Building on this principle, we propose \ac{ced}, a lightweight, training-free mechanism that detects distance jumps indicating cluster exits and adapts neighborhood sizes on a per-sample basis.

\par
The main contributions of this work are:
\begin{itemize}
\item We identify a structural failure mode of \ac{ldn} arising from neighborhood expansion across cluster boundaries and motivate the need for adaptive neighborhood selection to preserve local density structure.
\item We propose \ac{ced}, a lightweight, training-free algorithm for adaptive neighborhood selection based on distance jumps.
\item We conduct experiments using five embedding models across five benchmark datasets, demonstrating improved robustness to neighborhood-size selection and consistent performance gains over fixed, small-neighborhood baselines.
\end{itemize}

\section{Local density-based anomaly score normalization}
\label{sec:ldn}

\Ac{ldn} has been shown to improve robustness to domain shifts in distance-based anomaly detection \cite{wilkinghoff2025keeping,wilkinghoff2025local}. The key idea is to normalize anomaly scores using statistics derived from a local neighborhood in the embedding space, thereby accounting for variations in local data density. More recently, \Ac{varmin} was proposed to further stabilize normalized scores \cite{matsumoto2025adjusting}.

We briefly recap \ac{ldn}, its variance-minimized variant, and the role of the neighborhood size.

\subsection{Base anomaly scoring}
Let $\mathcal{X}_{\text{test}}\subset\mathbb{R}^{D}$ denote the set of test samples and $\mathcal{X}_{\text{ref}}\subset \mathbb{R}^{D}$ a reference set of normal training samples in the embedding space. All training samples from both domains are used as reference samples. For a test sample $x\in \mathcal{X}_{\text{test}}$, the base anomaly score is computed as the distance to its closest reference sample,
\begin{equation}
\score(x,\mathcal{X}_{\text{ref}})
:=
\min_{y \in \mathcal{X}_{\text{ref}}}
\dist(x,y)
\in\mathbb{R}_+,
\end{equation}
where $\dist:\mathbb{R}^{D}\times\mathbb{R}^{D}\rightarrow\mathbb{R}_+$ denotes a distance measure such as the Euclidean distance or cosine distance. 

\subsection{Local density-based normalization}
To mitigate performance degradation caused by domain shifts, \ac{ldn} normalizes the raw anomaly score using distances within a local neighborhood of each reference sample. For a given $y\in\mathcal{X}_{\text{ref}}$, a neighborhood of size $K\in\mathbb{N}$ is defined by its $K$ nearest neighbors in the reference set. Increasing $K$ implicitly assumes that locality is preserved as the neighborhood expands.

Combined with \ac{varmin} in log-space, this yields scaled anomaly scores of the form
\begin{equation}
\begin{aligned}
\score_{\text{scaled}}(x,\mathcal{X}_{\text{ref}}\mid K,\alpha)
&:= \min_{y\in\mathcal{X}_{\text{ref}}}
\bigl(
\log\dist(x,y) \\
&\hphantom{:={}}\;-\alpha\log\frac{1}{K}\sum_{k=1}^K\dist(y,y_{k})
\bigr)
\in \mathbb{R},
\end{aligned}
\end{equation}
where $y_k$ denotes the $k$-th nearest neighbor of $y$ in $\mathcal{X}_{\text{ref}}$. Setting $\alpha=1$ recovers the standard \ac{ldn} formulation.

\par
For \ac{varmin}, $\alpha=\alpha^*$ is selected to minimize the variance of the normalized scores over the reference set, i.e.,
\begin{equation}
\alpha^*
=
\arg\min_{\alpha \in \mathbb{R}}
\operatorname{Var}_{z \sim \mathcal{X}_{\text{ref}}}
\left(
\score_{\text{scaled}}\left(z, \mathcal{X}_{\text{ref}} \mid K, \alpha\right)
\right)
\in \mathbb{R}.
\end{equation}
This scoring backend is entirely training-free, requires no labels, and introduces no additional assumptions on the data distribution. Since the normalization constants depend only on the reference samples, they can be pre-computed without additional inference-time overhead.

\section{Locality violations and cluster exits}
\label{sec:cluster_exits}
\begin{figure}
    \centering
    \begin{adjustbox}{max width=\columnwidth}
          \begin{tikzpicture}
\begin{groupplot}[
    group style={
    group size=2 by 1,
    horizontal sep=2cm,},
	axis y line*=left,
    axis x line*=bottom,
    xmin=1,
    xmax=80,
    height=5cm,
    width=6cm,
    xticklabel style={align=center},
    yticklabel style={align=center},
    typeset ticklabels with strut,
    xlabel near ticks,
    ylabel near ticks,
    yticklabel style={xshift=-0.2cm},
    nodes near coords style={/pgf/number format/.cd,fixed zerofill,precision=2},
    ymajorgrids,
    scaled ticks = false,
    scaled y ticks = false,
    cycle list/Dark2
]
\nextgroupplot[title=(a) sorted distances, xlabel= neighbor index k, ylabel=distance $d_k$, legend style={at={(1.1,1.515)},anchor=north,legend columns=3,/tikz/every even column/.append style={column sep=0.5cm}},ymin=0.7,ymax=1.4]
\legend{source domain, target domain}
\addplot+[mark=triangle*,mark size=1,line width=1pt] coordinates {(1,0.787362277507782)(2,0.7996241450309753)(3,0.8074873089790344)(4,0.8139672875404358)(5,0.8190845847129822)(6,0.8237583637237549)(7,0.8280293345451355)(8,0.8323553204536438)(9,0.8368203639984131)(10,0.8407897353172302)(11,0.8449711799621582)(12,0.8491641283035278)(13,0.8533611297607422)(14,0.8573876023292542)(15,0.8617497682571411)(16,0.8667121529579163)(17,0.8717235922813416)(18,0.8774299621582031)(19,0.8835392594337463)(20,0.8947679996490479)(21,0.9096027612686157)(22,0.9164339900016785)(23,0.9233891367912292)(24,0.9355993866920471)(25,0.9715771079063416)(26,0.9781774282455444)(27,0.9833491444587708)(28,0.9875444173812866)(29,0.9923493266105652)(30,0.9953098893165588)(31,0.9996930956840515)(32,1.0026226043701172)(33,1.005387544631958)(34,1.008327841758728)(35,1.0117051601409912)(36,1.014829158782959)(37,1.0176903009414673)(38,1.0205397605895996)(39,1.0233955383300781)(40,1.0261991024017334)(41,1.0291637182235718)(42,1.0323153734207153)(43,1.0353708267211914)(44,1.0395290851593018)(45,1.0607513189315796)(46,1.0661357641220093)(47,1.0713071823120117)(48,1.0750426054000854)(49,1.0780137777328491)(50,1.090622067451477)(51,1.093281626701355)(52,1.095546841621399)(53,1.0976266860961914)(54,1.0996360778808594)(55,1.1013902425765991)(56,1.1031897068023682)(57,1.1050137281417847)(58,1.106794834136963)(59,1.1085915565490723)(60,1.1102137565612793)(61,1.111985206604004)(62,1.1139357089996338)(63,1.1156316995620728)(64,1.1172866821289062)(65,1.1191004514694214)(66,1.1208049058914185)(67,1.122503638267517)(68,1.1240922212600708)(69,1.1257331371307373)(70,1.127137541770935)(71,1.1285141706466675)(72,1.1298543214797974)(73,1.131272315979004)(74,1.1326632499694824)(75,1.134040355682373)(76,1.1353141069412231)(77,1.1365764141082764)(78,1.137890338897705)(79,1.1391758918762207)(80,1.140537977218628)};
\addplot+[mark=square*,mark size=1,line width=1pt] coordinates {(1,1.1585006713867188)(2,1.2145591974258423)(3,1.234057068824768)(4,1.2410573959350586)(5,1.2457278966903687)(6,1.2488155364990234)(7,1.2507076263427734)(8,1.2532408237457275)(9,1.2560430765151978)(10,1.2594496011734009)(11,1.2609000205993652)(12,1.2626594305038452)(13,1.2643457651138306)(14,1.2670392990112305)(15,1.2685327529907227)(16,1.26961088180542)(17,1.2706191539764404)(18,1.2720310688018799)(19,1.2738711833953857)(20,1.275423526763916)(21,1.2770105600357056)(22,1.2788636684417725)(23,1.2799487113952637)(24,1.2805678844451904)(25,1.2813814878463745)(26,1.28352689743042)(27,1.284600853919983)(28,1.2858525514602661)(29,1.287331461906433)(30,1.2886621952056885)(31,1.2896333932876587)(32,1.2908470630645752)(33,1.2916444540023804)(34,1.2923405170440674)(35,1.2931435108184814)(36,1.2945204973220825)(37,1.2954493761062622)(38,1.296120047569275)(39,1.2973400354385376)(40,1.2982368469238281)(41,1.2991359233856201)(42,1.3000208139419556)(43,1.3012030124664307)(44,1.3021057844161987)(45,1.3026612997055054)(46,1.3032190799713135)(47,1.303733229637146)(48,1.3045543432235718)(49,1.3050662279129028)(50,1.3058421611785889)(51,1.3065558671951294)(52,1.3072096109390259)(53,1.3079822063446045)(54,1.308416485786438)(55,1.3091142177581787)(56,1.3098517656326294)(57,1.3104901313781738)(58,1.3109934329986572)(59,1.3116626739501953)(60,1.3123421669006348)(61,1.3128734827041626)(62,1.3132244348526)(63,1.3137142658233643)(64,1.3141834735870361)(65,1.3149915933609009)(66,1.3155173063278198)(67,1.315917730331421)(68,1.316279649734497)(69,1.317193865776062)(70,1.317856788635254)(71,1.3182156085968018)(72,1.318682312965393)(73,1.3191524744033813)(74,1.319663405418396)(75,1.3201347589492798)(76,1.32095205783844)(77,1.3211838006973267)(78,1.3218462467193604)(79,1.322361707687378)(80,1.3228676319122314)};

\nextgroupplot[title=(b) distance ratios, xlabel=ratio index k , ylabel={$r_{k}=d_k/d_{k+1}$},ymin=0.94,ymax=1]
\addplot+[mark=triangle*,mark size=1,line width=1pt] coordinates {(1,0.9847692251205444)(2,0.9903078675270081)(3,0.9921330213546753)(4,0.9938446879386902)(5,0.9943757653236389)(6,0.994897186756134)(7,0.9948639273643494)(8,0.9947524666786194)(9,0.9953483939170837)(10,0.9951903223991394)(11,0.9951649308204651)(12,0.9952036142349243)(13,0.9953988790512085)(14,0.9950451850891113)(15,0.9944059252738953)(16,0.9943833947181702)(17,0.9937131404876709)(18,0.9932112097740173)(19,0.988152265548706)(20,0.9863464832305908)(21,0.9927749037742615)(22,0.9926310181617737)(23,0.9876207113265991)(24,0.9671628475189209)(25,0.9934335947036743)(26,0.9949843883514404)(27,0.9959668517112732)(28,0.9955224394798279)(29,0.997077226638794)(30,0.9960635304450989)(31,0.9971514344215393)(32,0.9972847104072571)(33,0.9971663951873779)(34,0.9967712759971619)(35,0.9969723224639893)(36,0.9971935153007507)(37,0.9972127079963684)(38,0.9971948266029358)(39,0.9972536563873291)(40,0.9971186518669128)(41,0.9969296455383301)(42,0.9970226287841797)(43,0.9959661960601807)(44,0.9824097752571106)(45,0.9949930310249329)(46,0.9950884580612183)(47,0.9964941740036011)(48,0.9972155690193176)(49,0.9889680743217468)(50,0.997621476650238)(51,0.9979871511459351)(52,0.9981260895729065)(53,0.9982017278671265)(54,0.9984424114227295)(55,0.9983991384506226)(56,0.9983838200569153)(57,0.9984297156333923)(58,0.9984316825866699)(59,0.9985787272453308)(60,0.9984633922576904)(61,0.9983754754066467)(62,0.9985449910163879)(63,0.9985637664794922)(64,0.9984479546546936)(65,0.9985273480415344)(66,0.9985511302947998)(67,0.9986372590065002)(68,0.9985622763633728)(69,0.9987618923187256)(70,0.9987810850143433)(71,0.9988026022911072)(72,0.9987454414367676)(73,0.99876868724823)(74,0.9987785816192627)(75,0.9988683462142944)(76,0.9988686442375183)(77,0.9988349080085754)(78,0.9988602995872498)(79,0.9988002777099609)(80,0.9987595081329346)};
\addplot+[mark=square*,mark size=1,line width=1pt] coordinates {(1,0.9507216215133667)(2,0.9835190773010254)(3,0.9944173097610474)(4,0.9962902069091797)(5,0.9975651502609253)(6,0.9985746145248413)(7,0.9979435205459595)(8,0.9977142214775085)(9,0.9973763227462769)(10,0.9988679885864258)(11,0.9986878633499146)(12,0.9986569285392761)(13,0.9979329109191895)(14,0.9988793134689331)(15,0.9991613626480103)(16,0.9991758465766907)(17,0.9989258050918579)(18,0.9986212849617004)(19,0.9988594055175781)(20,0.9988125562667847)(21,0.9986551403999329)(22,0.9991582632064819)(23,0.9995118975639343)(24,0.9993644952774048)(25,0.9985395669937134)(26,0.9991752505302429)(27,0.999071478843689)(28,0.9988666772842407)(29,0.9989674687385559)(30,0.9992908239364624)(31,0.9991286396980286)(32,0.9993739128112793)(33,0.9994441866874695)(34,0.9993659257888794)(35,0.9989539384841919)(36,0.9993308782577515)(37,0.9994617700576782)(38,0.999085545539856)(39,0.9993201494216919)(40,0.9992889165878296)(41,0.9993057250976562)(42,0.9990976452827454)(43,0.9993443489074707)(44,0.9995778203010559)(45,0.9995614886283875)(46,0.9995945692062378)(47,0.9993780851364136)(48,0.9995920062065125)(49,0.9994204640388489)(50,0.9994573593139648)(51,0.9994854927062988)(52,0.9993796348571777)(53,0.9996486902236938)(54,0.9995061755180359)(55,0.9994156956672668)(56,0.9995180368423462)(57,0.9996170997619629)(58,0.9995233416557312)(59,0.9994826316833496)(60,0.9995995759963989)(61,0.9997445940971375)(62,0.9996150732040405)(63,0.9996410608291626)(64,0.9994083642959595)(65,0.9995993375778198)(66,0.9996942281723022)(67,0.9997314214706421)(68,0.9993153810501099)(69,0.9995497465133667)(70,0.999714732170105)(71,0.9996346235275269)(72,0.9996511340141296)(73,0.9996291995048523)(74,0.9996454119682312)(75,0.999382495880127)(76,0.9998236894607544)(77,0.9994779825210571)(78,0.9996183514595032)(79,0.9996172189712524)(80,0.9997000694274902)};
\end{groupplot}
\end{tikzpicture}
    \end{adjustbox}
    \caption{Average sorted distances (left) and distance ratios (right) for BEATs embeddings of the \enquote{ToyCar} machine on the DCASE2025 dataset in the source and target domains. Pronounced distance jumps and low ratios mark cluster exits, which occur earlier in the target domain and reveal violations of locality under fixed neighborhood sizes.}
    \label{fig:motivation}
\end{figure}
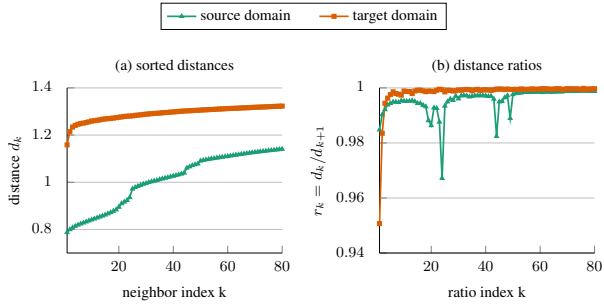

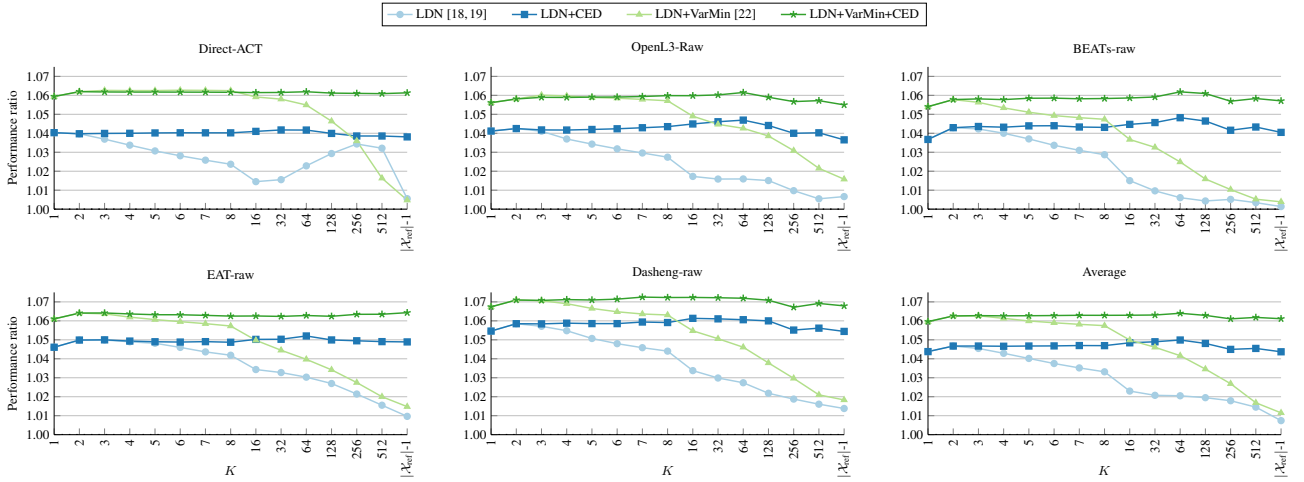
\begin{figure*}
    \centering
    \begin{adjustbox}{max width=\textwidth}
          \begin{tikzpicture}
\begin{groupplot}[
    group style={
    group size=3 by 2,
    xlabels at=edge bottom,
    ylabels at=edge left,
    horizontal sep=2cm,vertical sep=2cm,},
	axis y line*=left,
    axis x line*=bottom,
    ymin=0,
    ymax=7.5,
    legend style={at={(1.7,1.45)},anchor=north,legend columns=4,/tikz/every even column/.append style={column sep=0.5cm}},
    ylabel=Relative performance,
    xlabel=$K$,
    xlabel near ticks,
    xlabel style={yshift=3.5ex},
    height=5cm,
    width=10cm,
    ytick={-2,-1,0,1,2,3,4,5,6,7},
    yticklabels={0.98,0.99,1.00,1.01,1.02,1.03,1.04,1.05,1.06,1.07},
    xticklabel style={align=center,rotate=90},
    yticklabel style={align=center},
    xtick={1,2,3,4,5,6,7,8,9,10,11,12,13,14,15},
    xticklabels={1,2,3,4,5,6,7,8,16,32,64,128,256,512,$\lvert\mathcal{X}_\text{ref}\rvert$-1},
    xmin=1,
    xmax=15,
    ymajorgrids,
    cycle list/Paired
]
\nextgroupplot[title={Direct-\acs{act}}]
\legend{\acs{ldn} \cite{wilkinghoff2025keeping,wilkinghoff2025local}, \acs{ldn}+\acs{ced}, \acs{ldn}+\acs{varmin} \cite{matsumoto2025adjusting}, \acs{ldn}+\acs{varmin}+\acs{ced}}
\addplot+[mark=*,line width=1pt] coordinates {
(1,4.032)(2,3.974)(3,3.69)(4,3.374)(5,3.07)(6,2.814)(7,2.583)(8,2.368)(9,1.451)(10,1.555)(11,2.283)(12,2.932)(13,3.436)(14,3.213)(15,0.554)};

\addplot+[mark=square*,line width=1pt] coordinates {
(1,4.032)(2,3.974)(3,3.992)(4,3.999)(5,4.018)(6,4.026)(7,4.024)(8,4.023)(9,4.1)(10,4.175)(11,4.17)(12,3.993)(13,3.857)(14,3.856)(15,3.81)};

\addplot+[mark=triangle*,line width=1pt] coordinates {
(1,5.938)(2,6.196)(3,6.261)(4,6.251)(5,6.252)(6,6.27)(7,6.262)(8,6.247)(9,5.914)(10,5.8)(11,5.491)(12,4.637)(13,3.595)(14,1.632)(15,0.476)};

\addplot+[mark=star, mark size=2.5,line width=1pt] coordinates {
(1,5.938)(2,6.194)(3,6.178)(4,6.172)(5,6.173)(6,6.17)(7,6.167)(8,6.161)(9,6.145)(10,6.152)(11,6.19)(12,6.118)(13,6.104)(14,6.09)(15,6.132)};

\addplot[mark=none,line width=1pt,loosely dotted] coordinates {(1,0)(15,0)};

\nextgroupplot[title={OpenL3-Raw}]
\addplot+[mark=*,line width=1pt] coordinates {
(1,4.115)(2,4.248)(3,4.109)(4,3.699)(5,3.43)(6,3.179)(7,2.961)(8,2.74)(9,1.726)(10,1.587)(11,1.593)(12,1.508)(13,0.975)(14,0.546)(15,0.663)};

\addplot+[mark=square*,line width=1pt] coordinates {
(1,4.117)(2,4.248)(3,4.177)(4,4.171)(5,4.201)(6,4.237)(7,4.292)(8,4.35)(9,4.491)(10,4.608)(11,4.694)(12,4.411)(13,4.001)(14,4.025)(15,3.654)};

\addplot+[mark=triangle*,line width=1pt] coordinates {
(1,5.613)(2,5.81)(3,6.011)(4,5.968)(5,5.887)(6,5.851)(7,5.786)(8,5.713)(9,4.901)(10,4.47)(11,4.262)(12,3.857)(13,3.086)(14,2.152)(15,1.579)};

\addplot+[mark=star, mark size=2.5,line width=1pt] coordinates {
(1,5.613)(2,5.809)(3,5.895)(4,5.889)(5,5.908)(6,5.911)(7,5.939)(8,5.984)(9,5.981)(10,6.023)(11,6.15)(12,5.9)(13,5.67)(14,5.721)(15,5.497)};

\addplot[mark=none,line width=1pt,loosely dotted] coordinates {(1,0)(15,0)};

\nextgroupplot[title={BEATs-raw}]
\addplot+[mark=*,line width=1pt] coordinates {
(1,3.675)(2,4.297)(3,4.228)(4,4.002)(5,3.703)(6,3.367)(7,3.102)(8,2.869)(9,1.503)(10,0.965)(11,0.606)(12,0.432)(13,0.515)(14,0.343)(15,0.138)};

\addplot+[mark=square*,line width=1pt] coordinates {
(1,3.675)(2,4.295)(3,4.358)(4,4.321)(5,4.391)(6,4.398)(7,4.33)(8,4.311)(9,4.47)(10,4.565)(11,4.821)(12,4.646)(13,4.16)(14,4.33)(15,4.048)};

\addplot+[mark=triangle*,line width=1pt] coordinates {
(1,5.405)(2,5.774)(3,5.632)(4,5.34)(5,5.111)(6,4.932)(7,4.823)(8,4.738)(9,3.673)(10,3.258)(11,2.487)(12,1.591)(13,1.033)(14,0.525)(15,0.386)};

\addplot+[mark=star, mark size=2.5,line width=1pt] coordinates {
(1,5.403)(2,5.776)(3,5.812)(4,5.775)(5,5.847)(6,5.853)(7,5.819)(8,5.832)(9,5.861)(10,5.913)(11,6.184)(12,6.096)(13,5.691)(14,5.832)(15,5.709)};

\addplot[mark=none,line width=1pt,loosely dotted] coordinates {(1,0)(15,0)};

\nextgroupplot[title={EAT-raw}]
\addplot+[mark=*,line width=1pt] coordinates {
(1,4.611)(2,4.985)(3,4.991)(4,4.896)(5,4.818)(6,4.6)(7,4.363)(8,4.187)(9,3.433)(10,3.275)(11,3.031)(12,2.697)(13,2.142)(14,1.553)(15,0.965)};

\addplot+[mark=square*,line width=1pt] coordinates {
(1,4.609)(2,4.988)(3,4.998)(4,4.934)(5,4.898)(6,4.882)(7,4.904)(8,4.872)(9,5.023)(10,5.03)(11,5.203)(12,4.994)(13,4.949)(14,4.905)(15,4.89)};

\addplot+[mark=triangle*,line width=1pt] coordinates {
(1,6.101)(2,6.41)(3,6.369)(4,6.195)(5,6.067)(6,5.958)(7,5.852)(8,5.728)(9,4.974)(10,4.45)(11,3.971)(12,3.418)(13,2.746)(14,2.001)(15,1.477)};

\addplot+[mark=star, mark size=2.5,line width=1pt] coordinates {
(1,6.103)(2,6.409)(3,6.415)(4,6.364)(5,6.323)(6,6.321)(7,6.289)(8,6.25)(9,6.257)(10,6.234)(11,6.281)(12,6.24)(13,6.344)(14,6.348)(15,6.431)};

\addplot[mark=none,line width=1pt,loosely dotted] coordinates {(1,0)(15,0)};

\nextgroupplot[title={Dasheng-raw}]
\addplot+[mark=*,line width=1pt] coordinates {
(1,5.461)(2,5.849)(3,5.713)(4,5.48)(5,5.071)(6,4.797)(7,4.585)(8,4.404)(9,3.376)(10,2.987)(11,2.737)(12,2.182)(13,1.876)(14,1.607)(15,1.38)};

\addplot+[mark=square*,line width=1pt] coordinates {
(1,5.461)(2,5.849)(3,5.84)(4,5.878)(5,5.855)(6,5.854)(7,5.938)(8,5.911)(9,6.132)(10,6.105)(11,6.06)(12,5.999)(13,5.514)(14,5.615)(15,5.443)};

\addplot+[mark=triangle*,line width=1pt] coordinates {
(1,6.739)(2,7.101)(3,7.06)(4,6.905)(5,6.654)(6,6.478)(7,6.361)(8,6.309)(9,5.475)(10,5.058)(11,4.61)(12,3.782)(13,2.969)(14,2.099)(15,1.831)};

\addplot+[mark=star, mark size=2.5,line width=1pt] coordinates {
(1,6.738)(2,7.099)(3,7.078)(4,7.12)(5,7.1)(6,7.144)(7,7.252)(8,7.23)(9,7.235)(10,7.219)(11,7.192)(12,7.087)(13,6.718)(14,6.918)(15,6.789)};

\addplot[mark=none,line width=1pt,loosely dotted] coordinates {(1,0)(15,0)};

\nextgroupplot[title={Average}]
\addplot+[mark=*,line width=1pt] coordinates {
(1,4.379)(2,4.671)(3,4.546)(4,4.29)(5,4.018)(6,3.751)(7,3.519)(8,3.314)(9,2.298)(10,2.074)(11,2.05)(12,1.95)(13,1.789)(14,1.452)(15,0.74)};

\addplot+[mark=square*,line width=1pt] coordinates {
(1,4.379)(2,4.671)(3,4.673)(4,4.661)(5,4.673)(6,4.679)(7,4.698)(8,4.693)(9,4.843)(10,4.897)(11,4.99)(12,4.809)(13,4.496)(14,4.546)(15,4.369)};

\addplot+[mark=triangle*,line width=1pt] coordinates {
(1,5.959)(2,6.258)(3,6.267)(4,6.132)(5,5.994)(6,5.898)(7,5.817)(8,5.747)(9,4.987)(10,4.607)(11,4.164)(12,3.457)(13,2.686)(14,1.682)(15,1.15)};

\addplot+[mark=star, mark size=2.5,line width=1pt] coordinates {
(1,5.959)(2,6.257)(3,6.276)(4,6.264)(5,6.27)(6,6.28)(7,6.293)(8,6.291)(9,6.296)(10,6.308)(11,6.399)(12,6.288)(13,6.105)(14,6.182)(15,6.112)};

\addplot[mark=none,line width=1pt,loosely dotted] coordinates {(1,0)(15,0)};

\end{groupplot}
\end{tikzpicture}
    \end{adjustbox}
    \caption{Relative performance with respect to not applying \acs{ldn}, shown as a function of the neighborhood size $K$. Values correspond to the ratio between the performance obtained with the respective normalization method and that obtained without \acs{ldn}. Results are geometric means across all five evaluated datasets. For Direct-\acs{act}, results are averaged over ten independent trials.}
    \label{fig:scaling_ablation}
\end{figure*}

\Ac{ldn} assumes that nearby reference samples belong to the same local region in the embedding space.
Increasing the neighborhood size $K$ therefore assumes that this locality remains valid as the neighborhood expands.
In practice, \ac{ldn} is applied with very small neighborhoods, since increasing $K$ beyond one or two neighbors often leads to systematic performance degradation, even though larger neighborhoods are expected to yield more stable density estimates.
As shown in \cref{fig:motivation}(a), this degradation is structural rather than statistical.
Especially in the target domain, large distance jumps already occur among the first few neighbors.
Sorted distance profiles exhibit pronounced jumps when neighborhood expansion crosses cluster boundaries, directly indicating a violation of the locality assumption.
The resulting sensitivity to $K$ is further quantified in \cref{fig:scaling_ablation}.

\par
These jumps explain how locality is lost. For a given reference sample, distances typically increase smoothly within the same cluster.
When expansion crosses a cluster boundary, this smooth increase is interrupted by a sharp jump in distance. We refer to this transition as a \emph{cluster exit}.
As long as distances grow smoothly, neighborhood expansion remains valid. Once a cluster exit is encountered, further expansion includes samples from outside the local region, which corrupts local density estimates and degrades score normalization.
The core challenge is therefore not selecting a fixed neighborhood size, but identifying when locality no longer holds. 
In the following section, we present a training-free algorithm that detects cluster exits from distance jumps and uses them to adapt neighborhood sizes for local density estimation.
While developed for \ac{ldn}, the underlying principle is more general, and alternative mechanisms for detecting cluster exits could be integrated within the same framework.

\section{Cluster exit detection}

\Ac{ced} is a training-free mechanism for adaptive neighborhood selection in local density-based normalization.
Because local density estimates in \ac{ldn} are defined with respect to reference neighborhoods, all computations are carried out independently for each reference sample
$y \in \mathcal{X}_{\text{ref}}$ using only distances to other reference samples.
The method requires no labels or training, and can replace a fixed neighborhood size~$K$ in existing \ac{ldn} backends.
The central contribution is to demonstrate that adaptively selecting neighborhood sizes via cluster exit detection, rather than fixing $K$, is essential for preserving locality and stabilizing performance.
Fixed thresholds are used throughout all experiments.

\subsection{Distance ratios}

Let $K\geq2$ and let $\{y_k\}_{k=1}^{K}$ denote the $K$ nearest neighbors of a reference sample
$y \in \mathcal{X}_{\text{ref}}$, ordered by increasing distance. Define
\begin{equation}
    d_k(y) := \dist(y,y_k), \quad k=1,\dots,K .
\end{equation}
Within a local cluster, distances increase smoothly, whereas crossing a cluster boundary induces a sharp increase.
We capture such changes using distance ratios
\begin{equation}
    r_k(y) = \frac{d_k(y)}{d_{k+1}(y)+\epsilon}, \quad k=1,\dots,K-1,
\end{equation}
with $\epsilon=10^{-12}$. Since distances are ordered, $0<r_k(y)\leq1$, and small values indicate potential cluster exits.
For $K>2$, adjacent ratios are averaged,
\begin{equation}
    \tilde{r}_k(y)=\tfrac{1}{2}(r_k(y)+r_{k+1}(y)), \quad k=1,\dots,K-2,
\end{equation}
to reduce sensitivity to isolated fluctuations.
For $K=2$, no smoothing is applied.
The corresponding distance ratios are shown in \cref{fig:motivation}(b). Pronounced drops in the ratio sequence indicate potential cluster exits.

\subsection{Detecting cluster exits}

As the neighborhood size $k$ increases, a cluster exit is reflected by a pronounced drop in the smoothed ratio sequence $\tilde r_k(y)$. 
We therefore first identify the most informative transition by locating the index of the smallest ratio value,
\begin{equation}
k_{\mathrm{min}}(y)=\operatorname*{arg\,min}_{k\in\{1,\dots,K-2\}} \tilde{r}_k(y) .
\end{equation}
In practice, neighborhood expansion may already become unreliable at earlier indices if $\tilde r_k(y)$ drops below a conservative, data-adaptive threshold. 
To account for this, we define
\begin{equation}
\begin{aligned}
\mathcal{C}(y)=
\Bigl\{
&k\in\{1,\dots,K-2\} :
&\tilde{r}_k(y)
< Q_{0.04}(\tilde{\mathbf r}(y))
\Bigr\}
\end{aligned}
\end{equation}
where $Q_{0.04}(\tilde{\mathbf r}(y))$ denotes the $4$th percentile of the ratio sequence $\tilde{\mathbf r}(y)$.
If $\mathcal{C}(y)$ is nonempty, neighborhood expansion is truncated at the earliest candidate,
\begin{equation}
k_{\mathrm{ext}}(y)=\min \mathcal{C}(y),
\end{equation}
otherwise we set $k_{\mathrm{ext}}(y)=K-2$.
The adaptive neighborhood size is then chosen as
\begin{equation}
\hat{K}(y)=\min\{k_{\mathrm{ext}}(y),\,k_{\mathrm{min}}(y)\}+1.
\end{equation}
Since ratios require neighbor pairs, the index is effectively shifted, increasing the adaptive neighborhood size by one.

\subsection{Conservative fallback for sparse regions}

To avoid unreliable truncation in weakly structured regions, we first summarize the overall behavior of the ratio sequence by
$r_{\min}(y)=\min_k r_k(y)$.
Very strong ratio drops indicate that the reference point 
$y$ is located in a highly sparse region.
In such cases, we cautiously fall back to two neighbors when
\begin{equation}
r_1(y) < 0.85 \quad \text{or} \quad 
\frac{r_1(y)}{r_{\min}(y)} > 1.02 .
\end{equation}
We use two neighbors instead of one for consistency with the baseline \ac{ldn} without \ac{ced}.

\subsection{Integration into local density normalization}

Given the adaptive neighborhood size $\hat{K}(y)$, the local density estimate is computed as
\begin{equation}
    \mu(y)=\frac{1}{\hat{K}(y)}\sum_{k=1}^{\hat{K}(y)}\dist(y,y_k),
\end{equation}
and replacing the fixed neighborhood size in \ac{ldn} yields the final scaled score
\begin{equation}
\score_{\text{scaled}}^{\text{\ac{ced}}}(x,\mathcal{X}_{\text{ref}}\mid\alpha)
=
\min_{y\in\mathcal{X}_{\text{ref}}}
\bigl(\log\dist(x,y)-\alpha\log\mu(y)\bigr).
\end{equation}

\begin{table*}[!t]
\centering
\caption{Average performance for different normalization approaches across the development and evaluation sets of the DCASE2020, DCASE2022, DCASE2023,
DCASE2024, and DCASE2025 ASD datasets. $\Delta$ denotes improvement over baseline. CIs are 95\% paired-bootstrap intervals. Highest numbers in each column are in bold.}

\begin{adjustbox}{max width=\textwidth,max totalheight=\textheight}
\begin{NiceTabular}{cc *{6}{cc}}
\toprule
& & \multicolumn{10}{c}{\textbf{Embedding Model}} \\
\cmidrule{3-12}
&& \multicolumn{2}{c}{Direct-\acs{act}}& \multicolumn{2}{c}{OpenL3} & \multicolumn{2}{c}{BEATs} & \multicolumn{2}{c}{EAT} & \multicolumn{2}{c}{Dasheng} & \multicolumn{2}{c}{\textbf{Average Performance}}\\
\cmidrule(lr){3-4}\cmidrule(lr){5-6}\cmidrule(lr){7-8}\cmidrule(lr){9-10}\cmidrule(lr){11-12}\cmidrule(lr){13-14}
\textbf{Normalization}&\textbf{K}& average & {\scriptsize$\Delta$(CI)}& average & {\scriptsize$\Delta$(CI)}& average & {\scriptsize$\Delta$(CI)}& average & {\scriptsize$\Delta$(CI)}& average & {\scriptsize$\Delta$(CI)}& average & {\scriptsize$\Delta$(CI)}\\
\midrule
-&-&
$65.58\%$&{\color{gray}\scriptsize -}&
$61.77\%$&{\color{gray}\scriptsize -}&
$65.02\%$&{\color{gray}\scriptsize -}&
$61.82\%$&{\color{gray}\scriptsize -}&
$60.27\%$&{\color{gray}\scriptsize -}&
$62.90\%$&{\color{gray}\scriptsize -}\\
\midrule
\acs{ldn}&2&
$67.79\%$&{\color{gray}\scriptsize baseline}&
$64.62\%$&{\color{gray}\scriptsize baseline}&
$68.04\%$&{\color{gray}\scriptsize baseline}&
$64.99\%$&{\color{gray}\scriptsize baseline}&
$64.07\%$&{\color{gray}\scriptsize baseline}&
$65.90\%$&{\color{gray}\scriptsize baseline}\\
\acs{ldn}+\acs{ced}&64&
$67.91\%$&{\color{gray}\scriptsize +0.12 [0.02, 0.24]}&
$64.88\%$&{\color{gray}\scriptsize +0.26 [0.05, 0.47]}&
$68.36\%$&{\color{gray}\scriptsize +0.32 [0.03, 0.64]}&
$65.14\%$&{\color{gray}\scriptsize +0.12 [-0.10, 0.34]}&
$64.18\%$&{\color{gray}\scriptsize +0.11 [-0.04, 0.27]}&
$66.09\%$&{\color{gray}\scriptsize +0.19 [0.03, 0.35]}\\
\midrule
\acs{ldn}+\acs{varmin}&2&
\pmb{$69.25\%$}&{\color{gray}\scriptsize baseline}&
$65.56\%$&{\color{gray}\scriptsize baseline}&
$68.95\%$&{\color{gray}\scriptsize baseline}&
\pmb{$65.88\%$}&{\color{gray}\scriptsize baseline}&
$64.81\%$&{\color{gray}\scriptsize baseline}&
$66.89\%$&{\color{gray}\scriptsize baseline}\\
\acs{ldn}+\acs{varmin}+\acs{ced}&64&
$69.24\%$&{\color{gray}\scriptsize -0.01 [-0.07, 0.08]}&
\pmb{$65.75\%$}&{\color{gray}\scriptsize +0.19 [-0.01, 0.42]}&
\pmb{$69.20\%$}&{\color{gray}\scriptsize +0.25 [-0.01, 0.53]}&
$65.8\%$&{\color{gray}\scriptsize -0.08 [-0.17, 0.00]}&
\pmb{$64.86\%$}&{\color{gray}\scriptsize +0.05 [-0.10, 0.19]}&
\pmb{$66.96\%$}&{\color{gray}\scriptsize +0.07 [-0.04, 0.20]}\\
\bottomrule
\end{NiceTabular}
\end{adjustbox}
\label{tab:perf}
\end{table*}
\begin{table}[t]
\centering
\caption{Per-dataset improvements of \acs{ced} over the respective baseline (cf.~\cref{tab:perf}). 
$\Delta_{\text{avg}}$ denotes the mean improvement across embeddings (each averaged over splits), 
$\Delta_{\text{max}}$ the largest single-split gain for any embedding, 
and \#Emb$\uparrow$ the number of embeddings (out of 5) with positive average performance gain.}
\begin{adjustbox}{max width=\columnwidth,max totalheight=\textheight}
\begin{NiceTabular}{c *{2}{ccc}}
\toprule
 & \multicolumn{3}{c}{\textbf{without \ac{varmin}}} 
 & \multicolumn{3}{c}{\textbf{with \ac{varmin}}} \\
\cmidrule(lr){2-4} \cmidrule(lr){5-7}
\textbf{Dataset} 
& $\Delta_{\text{avg}}$ & $\Delta_{\text{max}}$ & \#Emb$\uparrow$ 
& $\Delta_{\text{avg}}$ & $\Delta_{\text{max}}$ & \#Emb$\uparrow$ \\
\midrule
DCASE2020 & $-0.06\%$ & $+0.13\%$ & 2/5 
           & $-0.12\%$ & $+0.01\%$ & 1/5 \\
DCASE2022 & $+0.29\%$ & $+0.55\%$ & 5/5 
           & $+0.10\%$ & $+0.37\%$ & 4/5 \\
DCASE2023 & $+0.32\%$ & $\mathbf{+1.39\%}$ & 5/5 
           & $+0.19\%$ & $\mathbf{+1.12\%}$ & 4/5 \\
DCASE2024 & $+0.18\%$ & $+0.68\%$ & 4/5 
           & $+0.13\%$ & $+0.64\%$ & 4/5 \\
DCASE2025 & $+0.20\%$ & $+0.82\%$ & 4/5 
           & $+0.06\%$ & $+0.72\%$ & 4/5 \\
\bottomrule
\end{NiceTabular}
\end{adjustbox}
\label{tab:delta_dataset}
\end{table}

\section{Experimental setup}
\subsection{Datasets}
We evaluate performance on five publicly available datasets for semi-supervised acoustic anomaly detection. We consider the DCASE2020 dataset \cite{koizumi2020description}, constructed from MIMII \cite{purohit2019mimii} and ToyADMOS \cite{koizumi2019toyadmos}; DCASE2022 \cite{dohi2022description}, based on MIMII-DG \cite{dohi2022mimiidg} and ToyADMOS2 \cite{harada2021toyadmos2}; DCASE2023 \cite{dohi2023description}, extending MIMII-DG and ToyADMOS2+ \cite{harada2023toyadmos2+}; DCASE2024 \cite{nishida2024description}, incorporating MIMII-DG, ToyADMOS2\# \cite{niizumi2024toyadmos2sharp}, and recordings collected under the IMAD-DS setup \cite{albertini2024imadds}; and DCASE2025 \cite{nishida2025description}, consisting of MIMII-DG, ToyADMOS2025 \cite{harada2025toyadmos2025}, and additional IMAD-DS recordings.
All datasets address semi-supervised \ac{asd} for machine condition monitoring across multiple machine types. They are partitioned into development and evaluation sets, with training splits containing only normal samples and test splits including both normal and anomalous recordings. Except for DCASE2020, which has a single domain, the datasets are designed to assess domain generalization by providing $990$ and $10$ source and target training samples per machine, respectively. In the test sets, machine types are known and domains are balanced, but explicit domain labels are omitted.

\par
All experiments follow the official evaluation protocols. For DCASE2020, we report the arithmetic mean of the AUC and pAUC \cite{mcclish1989analyzing} with $p=0.1$. For the remaining datasets, we report the harmonic mean of domain-specific AUCs and the domain-independent pAUC.

\subsection{Embedding models and baselines}

We evaluate five embedding models spanning task-specific and large-scale pre-trained representations. We include the Direct-\acs{act} model \cite{wilkinghoff2025local}, based on \cite{wilkinghoff2024self} and trained with the AdaProj loss \cite{wilkinghoff2024adaproj} using a subspace dimension of $32$. It employs FFT- and STFT-based feature branches and is trained with an \ac{act} objective over combined machine ID and attribute classes, complemented by a self-supervised feature exchange task. 
We further include $512$-dimensional openL3 embeddings \cite{cramer2019look} pre-trained on environmental sounds, BEATs \cite{chen2023beats} trained for three iterations on AudioSet \cite{gemmeke2017audioset}, EAT \cite{chen2024eat} pre-trained for $20$ epochs on AudioSet, and the base Dasheng model \cite{dinkel2024dasheng}. Following \cite{wilkinghoff2026temporal}, embeddings are aggregated using weighted generalized mean pooling with $p=3$ and relative deviation pooling weights ($\gamma=8$ for openL3, $\gamma=16$ for BEATs, $\gamma=1$ for EAT, and $\gamma=20$ for Dasheng). For EAT, embeddings are additionally pre-processed by thresholding low-valued components at $0.1$ and suppressing activation spikes via soft clipping using
$x \mapsto \tanh(x/0.5)\cdot 0.5$.

As baselines, we consider plain \ac{ldn} \cite{wilkinghoff2025keeping,wilkinghoff2025local} and \ac{ldn} combined with \ac{varmin} \cite{matsumoto2025adjusting}, following \Cref{sec:ldn}. Cosine distance is used for Direct-\acs{act}, while mean squared error is employed for all pre-trained embeddings.

\section{Results and discussion}

\subsection{Sensitivity to neighborhood size}
We first analyze the sensitivity of existing approaches, namely \ac{ldn} and \ac{ldn}+\ac{varmin}, to the choice of the neighborhood size.
As shown in \cref{fig:scaling_ablation}, average performance improves only when increasing the neighborhood size $K$ from $1$ to $2$.
This observation is consistent with the findings in \cite{wilkinghoff2025local}, which recommended using $K=1$ as a conservative default value for estimating the local neighborhood.
When increasing $K$ further, performance generally decreases monotonically and eventually approaches the level of not applying \ac{ldn}, except for Direct-\ac{act} without \ac{varmin}.
This behavior is observed for both \ac{ldn} with and without \ac{varmin} and verifies the claims made in \Cref{sec:cluster_exits}.

\subsection{Effect of cluster exit detection}

\Cref{fig:scaling_ablation} further illustrates the effect of cluster exit detection on the sensitivity of \ac{ldn} to the neighborhood size $K$. Across all embedding models, \ac{ced} markedly stabilizes performance for both plain \ac{ldn} and \ac{ldn}+\ac{varmin}. Whereas baseline performance typically peaks at $K\in{1,2}$ and degrades for larger neighborhoods, \ac{ced} enables robust performance across a wide range of $K$, with optimal values shifting to $K\in{16,32,64}$. This effect is more pronounced without \ac{varmin}. Since \ac{varmin} already compensates for part of the variability introduced by imperfect local-density estimates, the additional gains obtained from \ac{ced} are naturally smaller when both methods are combined. Large fixed neighborhoods alone do not yield similar gains, indicating that improvements stem from adaptive truncation rather than from increasing $K$ itself.

\par
Quantitative gains are summarized in \cref{tab:perf} and broken down per dataset in \cref{tab:delta_dataset}. Although global averages are moderate, improvements exhibit a clear structure across datasets. No systematic gains are observed on DCASE2020, consistent with its homogeneous data distribution and absence of explicit domain shifts, which reduce opportunities for pronounced local sub-cluster structure. In contrast, on DCASE2022–2025, \ac{ced} improves performance for at least 4 out of 5 embedding models, with average gains up to $+0.32\%$ and peak single-split improvements reaching $+1.39\%$ (DCASE2023). Without \ac{varmin}, improvements are frequently statistically significant. With \ac{varmin}, gains are smaller but remain structured on DCASE2022–2025, with peaks up to $+1.12\%$ (DCASE2023). Overall, these results indicate that adaptive locality preservation is particularly beneficial under heterogeneous local structure.

\section{Conclusion}
In this work, we analyzed the sensitivity of \ac{ldn} to neighborhood size and identified a failure mode that occurs when neighborhood expansion crosses cluster boundaries, thereby violating the locality assumption underlying density estimation.
Based on this insight, we proposed \ac{ced}, a lightweight mechanism that detects distance discontinuities to identify neighborhood exits and adapt score normalization accordingly.
Experiments across multiple embedding models and benchmark datasets showed that \ac{ced} reduces sensitivity to neighborhood-size selection and improves performance over a wide range of neighborhood sizes.
These results indicate that the observed performance degradation is not an inherent limitation of \ac{ldn}, but a consequence of silently violated locality assumptions.

\clearpage
\section{Generative AI disclosure}
Generative AI tools were used for language editing and polishing of the manuscript.
All scientific content, interpretations, and conclusions are the responsibility of the authors.

\bibliographystyle{IEEEtran}
\bibliography{refs}

\end{document}